\documentstyle[epsf]{mn}

\title[]{{\sl XMM-Newton} observations of the ultra-compact binary RX
J1914+24}

\author[]
{Gavin Ramsay$^{1}$, Pasi Hakala$^{2}$, Kinwah Wu$^{1}$, Mark Cropper$^{1}$, 
K. O. Mason$^{1}$ \and F. A.
C\'{o}rdova$^{3}$, W. Priedhorsky$^{4}$\\
$^{1}$Mullard Space Science Laboratory, University College London,
Holmbury St. Mary, Dorking, Surrey, RH5 6NT, UK\\
$^{2}$Observatory, University of Helsinki, PO Box 14,
FIN-00014 University of Helsinki, Finland\\
$^{3}$University of California, Riverside, CA 92521, USA\\
$^{4}$Los Alamos National Laboratory, MS D436, Los Alamos, NM 87545,
USA}

\begin{document}
\outer\def\gtae {$\buildrel {\lower3pt\hbox{$>$}} \over 
{\lower2pt\hbox{$\sim$}} $}
\outer\def\ltae {$\buildrel {\lower3pt\hbox{$<$}} \over 
{\lower2pt\hbox{$\sim$}} $}
\newcommand{\ergscm} {ergs s$^{-1}$ cm$^{-2}$}
\newcommand{\ergss} {ergs s$^{-1}$}
\newcommand{\ergsd} {ergs s$^{-1}$ $d^{2}_{100}$}
\newcommand{\pcmsq} {cm$^{-2}$}
\newcommand{\ros} {\sl ROSAT}
\newcommand{\exo} {\sl EXOSAT}
\newcommand{\asca} {\sl ASCA}
\newcommand{\xmm} {\sl XMM-Newton}
\newcommand{\chan} {\sl Chandra}
\def\rchi{{${\chi}_{\nu}^{2}$}}
\def\uchi{{${\chi}^{2}$}}
\newcommand{\Msun} {$M_{\odot}$}
\newcommand{\Mwd} {$M_{wd}$}
\def\Mdot{\hbox{$\dot M$}}
\def\mdot{\hbox{$\dot m$}}

\maketitle

\begin{abstract}

We present {\xmm} observations of the 569 sec period system RX
J1914+24 (V407 Vul). This period is believed to represent the binary
orbital period making it an ultra-compact binary system. By comparing
the phase of the rise to maximum X-ray flux at various epochs (this
includes observations made using {\ros}, {\asca} and {\chan}) we find
that the system is spinning up at a rate of
3.17$\pm0.07\times10^{-12}$ s/s. We find that the spectra softens as
the X-ray flux declines towards the off-phase of the 569 sec
period. Further, the spectra are best fitted by an absorbed blackbody
component together with a broad emission feature around 0.59keV. This
emission feature is most prominent at the peak of the on-phase. We
speculate on its origin.

\end{abstract}


\section{Introduction}

RX J1914+24 (also known as V407 Vul) is one of 3 sources discovered in
recent years which show intensity variations on periods of less than
$\sim$10 mins. As no other periods have been detected in these
systems, and for other reasons, these periods have been associated
with the binary orbital period. As such, these systems would have the
shortest binary period of any known system. In addition, they would be
amongst the strongest sources of constant gravitational radiation in
the sky and easily detectable using the future {\sl LISA} space
mission. Their nature, however, remains controversial.

Of the 3 systems, ES Cet (Warner \& Woudt 2002), with a period of 620
sec, has been shown to have an accretion disc. Both RX J1914+24
(Cropper et al 1998, Ramsay et al 2000, 2002) with a period of 569
sec, and RX J0806+15 (Ramsay, Hakala \& Cropper 2002, Israel et al
2002) with a period of 321 sec, do not show evidence for an accretion
disc and share many similar properties. Their X-ray light curves are
almost identical, being `off' for around half their cycle, showing a
sharp rise to maximum flux and a more gradual decay. In contrast,
their optical light curves are sinusoidal in shape, and in anti-phase
with the X-ray phase (Ramsay et al 2000, Israel et al 2003). The
period of both systems are reported to be evolving in the same
direction (ie spinning up) as predicted if their binary orbit is
evolving through gravitational radiation (Hakala et al 2003,
Strohmayer 2003, Hakala, Ramsay \& Byckling 2004 for RX J0806+15 and
Strohmayer 2002, 2004a for RX J1914+24).

They do, however, differ in some respects. RX J0806+24 shows weak
optical emission lines, with Hydrogen blending with Helium lines
(Israel et al 2002, Norton, Haswell \& Wynn 2004). On the other hand
RX J1914+24 shows a generally featureless optical spectrum but with
weak absorption lines which appear similar to that of a K star
(Steeghs et al 2004). At present it is unclear as to how to interpret
this spectrum, although a triple system is a possibility.

RX J1914+24 has been observed in X-rays using {\sl ROSAT}, {\sl ASCA}
(Cropper et al 1998, Ramsay et al 2000, 2002) and {\sl Chandra}
(Strohmayer 2004a). With its larger effective area, {\sl XMM-Newton}
provides the possibility of obtaining phase resolved spectroscopy
through the 569 sec cycle. Here, we present observations of RX
J1914+24 made using {\xmm}.

\section{Observations and Data Reduction}

{\xmm} was launched in Dec 1999 by the European Space Agency. The EPIC
instruments contain imaging detectors covering the energy range
0.15--10keV with moderate spectra resolution (EPIC pn, Str\"{u}der et
al 2001, EPIC MOS, Turner et al 2001), while the RGS has a high
spectral resolution between 0.3--2keV (den Herder et al 2001). The
observation log is shown in Table \ref{log}.

The EPIC pn detector was configured in small window mode and thin
filter, the EPIC MOS1 detector in timing mode and thin filter, and the
EPIC MOS2 detector in small window mode and medium filter. The Optical
Monitor used two UV filters (UVW2 and UVM2). Because of the high
extinction to RX J1914+24 (Cropper et al 1998), RX J1914+24 was not
detected, as expected, in either filter.

The X-ray data were processed using the {\sl XMM-Newton} {\sl Science
Analysis Software} (SAS) v6.0. The data were barycentrically corrected
and in units of TT. For the EPIC detectors (Str\"{u}der et al 2001,
Turner et al 2001), data were extracted using an aperture of 40$^{''}$
centered on the source position: this $\sim$87 percent of the
encircled energy. Background data were extracted from a source free
region. The background data were scaled and subtracted from the source
data. We show the mean background subtracted EPIC pn count rate for
the two observations in Table \ref{log}. We extracted the RGS spectra
in the standard way using {\tt rgsproc}. Although they were of low
signal-to-noise no bright distinct lines were detected.

\begin{table}
\begin{center}
\begin{tabular}{lrrr}
\hline
{\sl XMM} & Start Date & Duration & Mean EPIC\\
 Orbit   &            & (ksec)   & pn (Ct/s) \\
\hline
0718 & 2003-11-09:22:36:10 & 9.5 & 0.80 \\
0721 & 2003-11-15:22:08:26 & 8.5 & 0.73 \\
\hline
\end{tabular}
\end{center}
\caption{The observation log of {\xmm} observations of RX
J1914+24. The start time is in UTC.}
\label{log}
\end{table}

\section{Light Curves}

\begin{figure}
\begin{center}
\setlength{\unitlength}{1cm}
\begin{picture}(8,10.5)
\put(0.0,-1){\includegraphics{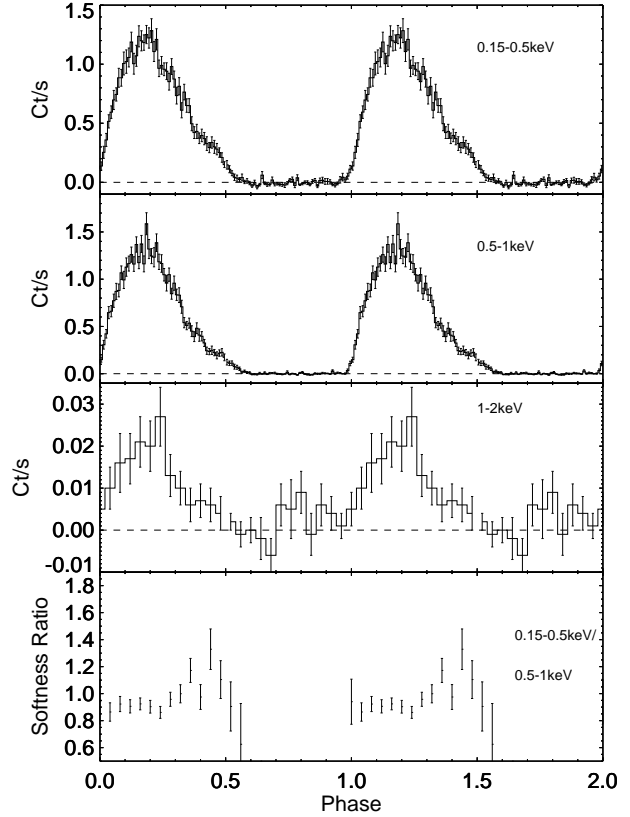}}
\end{picture}
\end{center}
\caption{From the top: The folded and binned light curve in the
0.15--0.5keV band, the 0.5--1.0keV band, the 1--2keV band, and the 
softness curve (0.15--0.5/0.5--1.0keV). The soft and medium bands are
binned into 100 bins, the hard band 25 bins and the softness ratio
curve 25 bins. The data have been folded on $T_{o}$=2449258.033373137
(BJD), and the mean period between $T_{o}$ and $T_{XMM}$ (to ensure
proper phasing of the data).}
\label{fold}
\end{figure}

We show the folded and binned light curve in 3 energy bands in Figure
\ref{fold}. As found by previous X-ray studies, the system is `off'
for $\sim$0.4 cycles. We extracted images of the field of RX J1914+24
during the faint phase using the EPIC pn data and do not detect RX
J1914+24. After the off-phase, there is a sharp increase in flux in
the 0.15--0.5keV (soft) and 0.5--1.0keV (medium) bands. The softness
curve (soft/medium), (Figure \ref{fold}), shows that the spectrum of
RX J1914+24 softens towards the descent from the peak intensity of the
bright phase.  This implies that the X-ray emission region has a
temperature structure. We also show the 1--2keV (hard) folded light
curve in Figure \ref{fold}: the count rate is much reduced compared to
the soft or medium bands. It shows a pronounced drop in intensity at
the point where the softness curve starts to soften.

Ramsay et al (2000) showed using {\sl ASCA} data that RX J1914+24 has
no significant flux above 2keV. With the higher effective area of
{\xmm} compared to {\sl ASCA} we can test this assertion more
rigorously. We extracted images of the immediate field around RX
J1914+24 using data from the EPIC pn detector taken at both
epochs. There is no significant detection above 2.0keV in either
epoch. To determine an upper limit to the hard X-ray emission we added
a Bremsstrahlung component to the model described in \S
\ref{spec}. For a temperature of $kT$=10keV we find an upper limit to
the 2-10keV unabsorbed flux of 1.3$\times10^{-13}$ \ergscm, which
reduces to 9.6$\times10^{-15}$ \ergscm for $kT$=1keV.

We searched for periods above 20 min using all the {\ros}, {\asca},
{\chan} and {\xmm} data, and just the {\xmm} separately. We found no
evidence for a significant period(s) in either combination of data.
(Care has to be taken since the {\ros} data are in units of UTC;
{\asca} is on a system very close to UTC, while {\chan} and {\xmm}
data are on TT).

\section{Phasing of the data}
\label{phase}

Strohmayer (2002) performed a coherent timing analysis of {\ros} and
{\asca} data and found that the 569 sec period in RX J1914+24 was
spinning up at a rate of 8$\pm3\times10^{-18}$ Hz/s
(=2.6$\times10^{-12}$ s/s). This was later refined with the addition
of {\chan} data to 7.0$\pm0.8\times10^{-18}$ Hz/s
(=2.3$\times10^{-12}$ s/s, Strohmayer 2004a). This result was
consistent with the expected spin up rate if the systems was being
driven entirely by gravitational radiation and the secondary star has
a low mass.

\begin{figure}
\begin{center}
\setlength{\unitlength}{1cm}
\begin{picture}(8,11.2)
\put(-1.,-0.7){\includegraphics{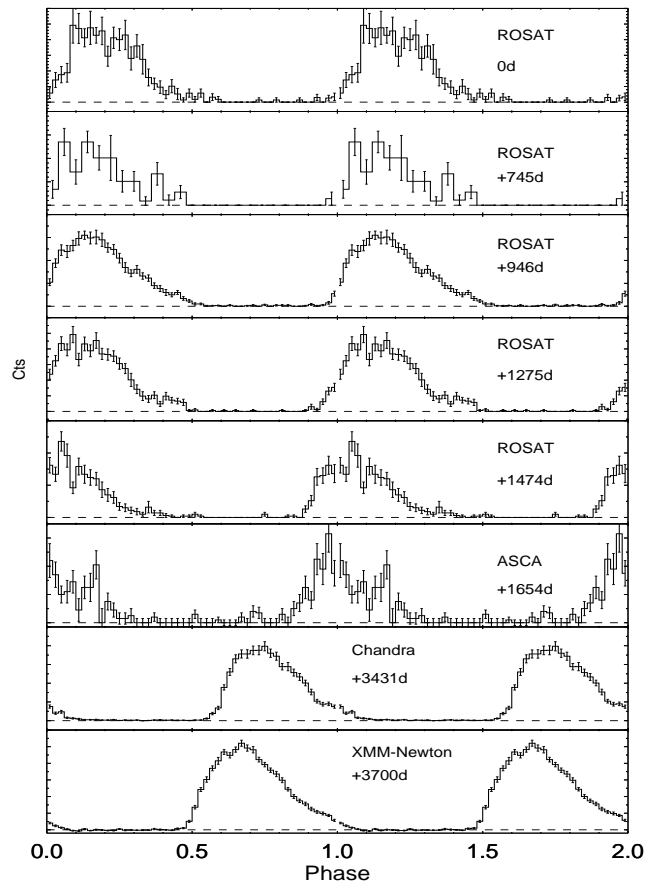}}
\end{picture}
\end{center}
\caption{The {\ros}, {\asca}, {\sl Chandra} and {\xmm} data folded on
the best fit period of Strohmayer (2004a). It shows the phase at which
the rise to maximum in X-rays occurs earlier over time - ie it is
spinning up.}
\label{period}
\end{figure}

The spin-up in both RX J0806+15 and RX J1914+24 has been met with some
degree of scepticism (eg Woudt \& Warner 2003). To show the effect of
the spin-up in RX J1914+24 we show in Figure \ref{period} the {\ros},
{\asca}, {\chan} (which were extracted from the {\chan} archive and
reduced in the same manner as Strohmayer 2004a) and {\xmm} data folded
on the constant period term (1/$\nu_{o}$) of Strohmayer (2004a). The
phase of the rise to maximum flux has continued to arrive
progressively earlier since the epoch of the {\chan} data: it is clear
that RX J1914+24 is spinning up. It is highly unlikely that the true
period is a side-band power peak Strohmayer (2004a).

\begin{figure}
\begin{center}
\setlength{\unitlength}{1cm}
\begin{picture}(8,7.8)
\put(-0.5,-4.){\includegraphics{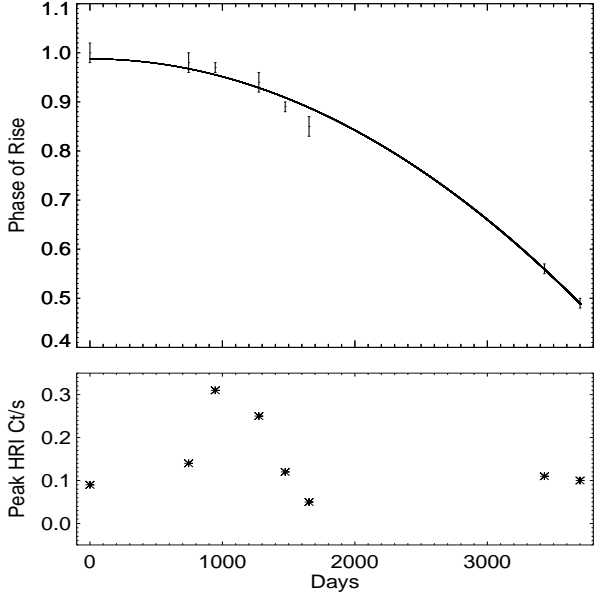}}
\end{picture}
\end{center}
\caption{Top panel: The phase of the sharp rise to maximum X-ray flux
determined for all the epochs shown in Figure \ref{period} determined
using a constant period, $P_{o}$. The solid line shows the best fit to
the data giving a spin-up rate of 3.17$\times10^{-12}$ s/s.  Lower
Panel: The X-ray intensity of the peak of the bright phase expressed
in equivalent {\ros} HRI Ct/s. The start date is JD=2449259. We show
the error on the phase of the sharp rise, while the error on the X-ray
intensity is smaller than the plotted symbol.}
\label{pdot}
\end{figure}

For the phased light curves shown in Figure \ref{period} we estimated
the start of the bright phase by noting the phase at which the count
rate rose by a significant level compared to the previous phase
bin. We show how the start of the bright phase varies over time in
Figure \ref{pdot} together with an estimate of the uncertainty in the
phase. We fitted these points with a constant $\dot{P}$ term: the best
fit is shown as a solid line in Figure \ref{period}. We note that the
shift in phase for a constant $\dot{P}$ term is not linear over time
(see Cropper et al 2004). We find a spin-up rate of
3.17$\pm0.10\times10^{-12}$ s/s (=1.0$\times10^{-17}$ Hz/s). Because
the errors on the phase of the rise to the bright phase are not
strictly 1$\sigma$ errors, we determined the error using a bootstrap
method. This gave a larger error than the formal error to the fit
(0.07$\times10^{-12}$ s/s). The spin-up rate which we determine is
slightly greater than the spin-up rate derived by Strohmayer (2004a).

We also show the long term X-ray intensity of RX J1914+24 in the lower
panel of Figure \ref{pdot}. These values were obtained by converting
the peak X-ray count rate in each epoch to the equivalent {\ros} HRI
count rate (since most observations were determined using this
instrument). We used the spectral parameters shown Table \ref{fits}
and the {\tt PIMMS} tool (Mukai 1993) to convert the count rate in
other detectors to that expected for the {\ros} HRI. We note that the
largest residual to the fit occurred after the peak X-ray flux had
decreased from a short epoch of enhanced X-ray emission. However, the
deviation is small, (1.9$\sigma$), so therefore probably not
significant.

Ramsay et al (2000) showed that the peak of the X-ray and optical band
emission were offset in phase. Since their {\asca} and $I$ band data
were taken within 3 months of each other, this phase offset is still
valid despite the spin-up of the system.

\section{Spectra}
\label{spec}

\begin{figure}
\begin{center}
\setlength{\unitlength}{1cm}
\begin{picture}(8,5.5)
\put(-0.5,-0.3){\includegraphics{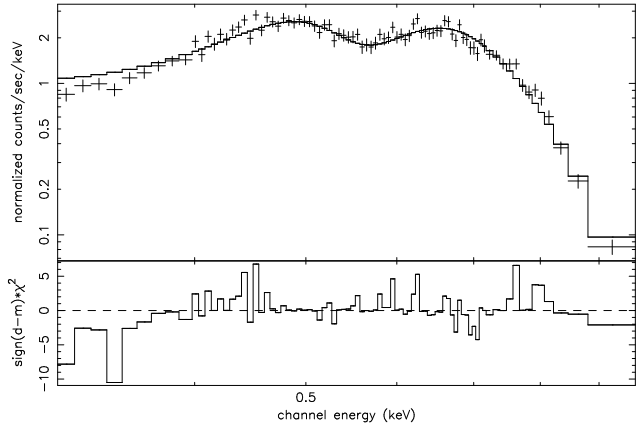}}
\end{picture}
\end{center}
\caption{The integrated EPIC pn spectrum together with the best fit
absorbed blackbody plus Gaussian line. We ignored events with energies
below 0.3keV and above 1keV and binned the spectrum so that there was
a minimum of 50 counts per min.}
\label{spec_fig}
\end{figure}

To fully utilise the data from the two observations (cf Table 1), we
combined the EPIC pn event files from the two epochs (we also did this
for EPIC MOS2). We used single and double events for the pn spectrum,
and single to quadruple events for the MOS spectrum, and those events
with FLAG=0. We created response and auxillary files using the SAS
tasks {\tt arfgen} and {\tt rmfgen}. The spectra are shown in Figure
\ref{spec_fig}. Ramsay et al (2000) fitted X-ray spectra taken using
{\ros} and {\asca} using an absorbed blackbody model. We used {\tt
XSPEC} (Arnaud 1996) to fit the spectra, binned them so there was a
minimum of 50 counts per min and we applied a low energy cut-off of
0.25keV. The {\xmm} spectra are not well fitted using such a
model. Various other models were fitted including; an absorbed two
temperature blackbody model; a thermal plasma model with varying
element abundances; the previous model together with a blackbody; and
an emission model together with a line in {\sl absorption} around
0.55keV. The only model which even approached a reasonable fit
(\rchi=1.63, 79 dof in EPIC pn) was an absorbed blackbody plus broad
Gaussian line in emission at 0.59 keV: this was an unexpected result
(the fit to the EPIC MOS2 spectra was poorer giving \rchi=2.20, 29 dof
-- the response of the MOS cameras has changed over time, especially
at energies $<$0.5keV). The spectral parameters for the integrated
EPIC pn spectrum together with the flux is shown in Table
\ref{fits}. Spectra were also extracted from both RGS instruments: no
distinct strong individual lines were detected with any confidence.

\begin{table}
\begin{center}
\begin{tabular}{lrrr}
\hline
N$_{H}$ & 4.2$\pm0.1\times10^{21}$ \pcmsq \\
$kT_{bb}$ & 58.6$^{+0.9}_{-1.2}$ eV\\
line center & 0.586$\pm0.004$keV\\
EW & 257$^{+23}_{-9}$ eV\\
Flux$^{o}$ & 1.35$^{+0.05}_{-0.04}\times10^{-12}$ \ergscm\\
Flux$^{u}$ & 3.18$^{+0.10}_{-0.16}\times10^{-10}$ \ergscm\\
\hline
\end{tabular}
\end{center}
\caption{The best fit spectral parameters to the integrated EPIC pn
spectrum. Flux$^{o}$ refers to the observed flux in the 0.1--10keV
energy band and Flux$^{u}$ refers to the unabsorbed, bolometric flux.}
\label{fits}
\end{table}

\begin{figure*}
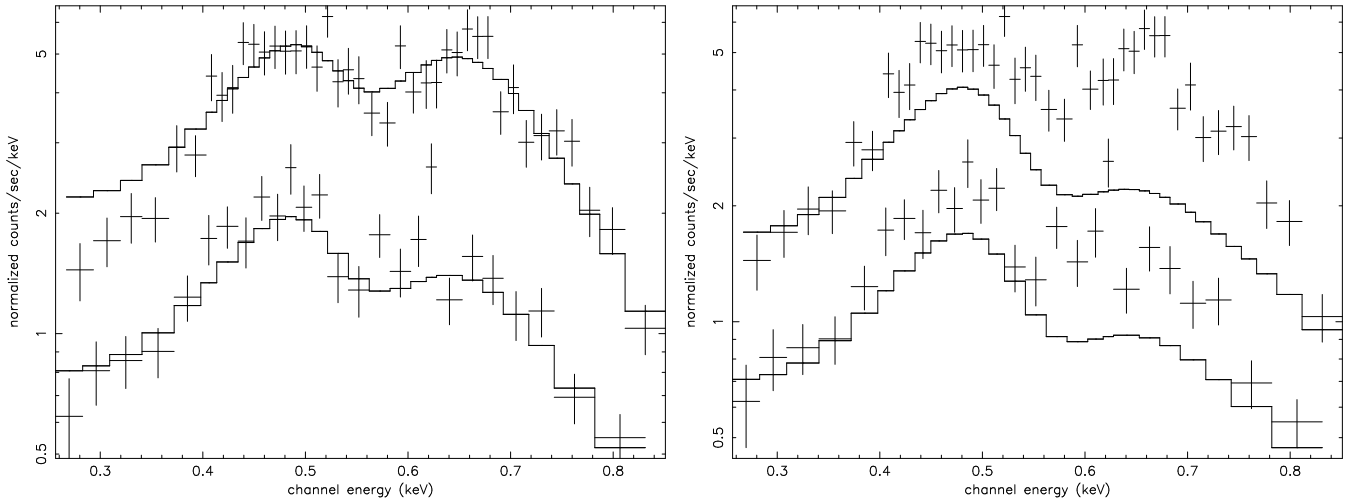

\begin{center}
\setlength{\unitlength}{1cm}
\begin{picture}(12,6.5)
\put(-3.5,-0.5){\includegraphics{phase_res_paper_withline_rev.ps}}
\put(5.5,-0.5){\includegraphics{phase_res_paper_noline_rev.ps}}
\end{picture}
\end{center}
\caption{In the left hand panel we show 2 of the phase resolved EPIC
pn spectra together with the best fit using an absorbed blackbody plus
Gaussian line near 0.59keV. In the right hand panel we emphasise the
effect that the line has on the fit by setting its normalisation to
zero. In each panel the upper spectrum covers $\phi$=0.1-0.25 and the
lower spectrum $\phi$=0.35--0.50. The spectra covering the other phase
intervals have been omitted for clarity.}
\label{spec_phase}
\end{figure*}

There are still significant residuals in the fit to the integrated
spectrum using the absorbed blackbody plus line model. This is not
unexpected since the softness ratio variation (Figure \ref{fold})
suggests that the shape of the spectrum changes over the bright X-ray
phase. We therefore extracted four spectra covering the bright phase
from the EPIC pn data. Again our model was an absorbed blackbody plus
Gaussian line in emission near 0.59keV. The spectral parameters for
each spectrum were tied together apart from the normalisations which
were allowed to vary. A simultaneous fit to the four spectra gave
\rchi=1.45 (170 dof): the fit was not improved by letting the
blackbody temperature vary. We find that the normalisation of both the
blackbody and line components roughly follow the shape of the X-ray
light curve, although the normalisation of the blackbody falls more
rapidly compared to that of the Gaussian after intensity maximum
resulting in a rise in the equivalent width. Although the Gaussian
line may not be physically realistic, it does, however, provide a
suitable way of characterising the spectra. To show the effect of the
Gaussian line, we show in Figure \ref{spec_phase} the fits to phase
resolved spectra including the line and then when we switch the line
normalisation to zero: the line has a very significant effect on the
fits.

\begin{figure}
\begin{center}
\setlength{\unitlength}{1cm}
\begin{picture}(8,11)
\put(0.5,0){\includegraphics{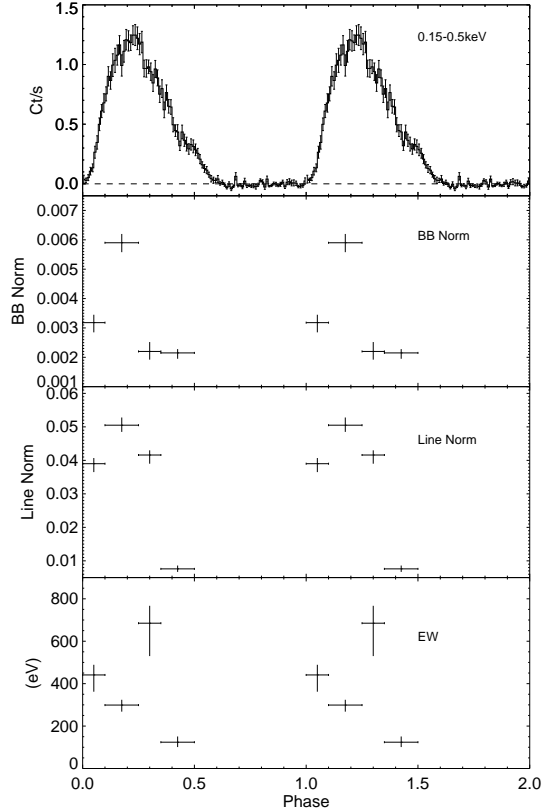}}
\end{picture}
\end{center}
\caption{From the top: the 0.15--0.5 keV light curve folded on the 569
sec and binned; the normalisation of the blackbody
component; the normalisation of the Gaussian line
included near 0.59keV; the equivalent width of the Gaussian line.}
\label{phase}
\end{figure}

In light of these results, we revisited the {\ros} PSPC integrated
spectrum and used the above model to fit that spectrum (extracted from
the public archive in `Rev2' calibrated form). We tied the
normalisation of the blackbody and Gaussian components so that they
were the same ratio as found in the integrated {\xmm} fits. We find
that this model gives a better fit (\rchi=1.18 6dof) compared to an
absorbed blackbody (\rchi=1.58 6dof). We note that the resulting fit
give a lower absorption column ($N_{H}=1.7\times10^{21}$ \pcmsq) and
higher blackbody temperature ($kT_{bb}$=78eV) compared to the fits of
Cropper et al (1998). Ramsay \& Cropper (2004) have noted that {\ros}
observations of magnetic cataclysmic variable stars have shown a
difference in the spectral parameters using `Rev0' calibrated data and
the `Rev2' calibrated data.

\section{Discussion}
\label{summary}

We have presented observations of the X-ray source RX J1914+24
obtained using {\xmm}. We have two main results: we find that RX
J1914+24 is spinning up at a rate slightly greater than that
determined by Strohmayer (2004a) and we find that the X-ray spectrum
is very unusual showing a broad emission line at $\sim$0.59keV in
addition to the expected soft blackbody component.

\subsection{The spin history of RX J1914+24}
\label{spin_history}

The models which have been used to explain the nature of RX J1914+24
and RX J0806+15 fall into two general categories: those involving
accretion and those which do not. At face value, the results which
show both RX J1914+24 and RX J0806+15 spinning up appear to rule out
most of the accretion driven models: the double degenerate polar model
(Cropper et al 1998) and the direct impact model (Marsh \& Steeghs
2002, Ramsay et al 2002). Unlike these models, in which the observed
intensity variation is attributed to the binary orbital period,
Norton, Haswell \& Wynn (2004) attribute it to the spin period of an
accreting white dwarf. No second period is detected since in this
model the binary inclination is close to face-on. 

The non-accreting model is the Unipolar-Inductor (UI) model proposed
by Wu et al (2002).  In this model, a non-magnetic white dwarf
traverses the magnetic field of the primary magnetic white dwarf,
which causes large currents to be driven. Resistive dissipation occurs
at the foot-points of the primary white dwarf and X-rays are
released. This model also predicts that the system is an
electron-cyclotron maser source in the UI phase (Willes, Wu \& Kuncic
2004).

It is important to note that these systems have been observed for less
than 10 years and therefore the period decrease may simply be due to a
slow variation rather than a long term trend.

In spite of the complicated interaction between the stellar spin rates
and the binary orbital period, the change in the binary orbital period
is determined only by the energy and angular momentum losses and
redistribution in the system. 

We illustrate this using the UI model given in Wu et al (2002).  For a
binary with a secondary star in synchronous rotation with the orbit,
it can be shown that the change in the orbital angular velocity
$\omega_{\rm o}$ is given by

\begin{equation} g(\omega_{\rm o}) \left({{\dot{\omega_{\rm
o}}}\over{\omega_{\rm o}}} \right) + {2\over 5 } M_1 R_1^2 \omega_1
{\dot{\omega_1}} \ = \ {\dot E}_{\rm gr} + {\dot E}_{\rm diss} \ ,
\end{equation} 

\noindent where ${\dot E}_{\rm gr}$ is the energy loss due to
gravitational radiation, ${\dot E}_{\rm diss}$ is the energy loss due
to dissipation, and $\omega_{1}$ is the angular velocity of the
primary (see Wu et al (2002) for the explicit expression of
$g(\omega_{\rm o})$).
            
In the UI model, the energy dissipation is caused by the resistive
heating at the foot-points of the magnetic field lines on the magnetic
white dwarf. In terms of the asynchronicity parameter of the spin of the
primary, $\alpha~(\equiv \omega_1/ \omega_{\rm o})$, the energy
dissipation

\begin{equation}
{\dot E}_{\rm diss} \ = \ -{2 \over 5}\left| (1- \alpha) M_1 R_1^2
\omega_{\rm o} {\dot \omega}_1 \right| . 
\end{equation} 

\noindent It can be shown that

\begin{equation} \left({{\dot{\omega_{\rm o}}}\over{\omega_{\rm o}}}
\right) \ = \ {1 \over {g(\omega_{\rm o})}} \left[ {\dot E}_{\rm gr}
+ {2\over 5}\alpha M_1 R_1^2 \omega_{\rm o}^2
\left({{\dot{\omega_1}}\over{\omega_1}} \right) \right] \ .
\end{equation} 

\noindent As $({\dot \omega}_1/\omega_1) = ({\dot \omega}_{\rm
o}/\omega_{\rm o}) + ({\dot \alpha}/\alpha)$, the rate of change in
the orbital angular velocity may be expressed as

\begin{equation}
\left({{\dot{\omega_{\rm o}}}\over{\omega_{\rm o}}} \right) \ = \
{{{\dot E}_{\rm gr}} \over {g(\omega_{\rm o})}} \left[ 1 + {2 \over
5}{{ \alpha M_1 R_1^2 {\omega_{\rm o}^2} } \over {{\dot E}_{\rm gr}}
} \left({{\dot \alpha}\over{\alpha}} \right) \right] \left[ 1 - {{2
\over 5} {{\alpha M_1 R_1^2 {\omega_{\rm o}}^2} \over {g(\omega_{\rm
o})}}}\right]^{-1} \ 
\end{equation} 

\noindent This implies that the secular evolution of the orbital
angular velocity is also determined by the coupling between the spin
of the primary and the orbital rotation in spite of the ultimate
driver being the energy and angular momentum losses due to
gravitational radiation. More detailed work is needed to determine 
the range of $\alpha$ which could give rise to the observed spin-up
rate. 

\subsection{X-ray luminosity}

The lower panel of Figure \ref{pdot} shows that the observed count
rate of RX J1914+24 varies by an order of magnitude. At the {\xmm}
epoch, the unabsorbed, bolometric flux determined using the model fits
is shown in Table \ref{fits}.  Due to projection effects we correct
the observed count rate by the ratio of the peak to mean count rate (a
factor of 3.4) which gives a luminosity of $\sim1\times10^{35}$ \ergss
(for 1kpc, Steeghs et al 2004). For the same distance this is $\sim$2
orders of magnitude lower than the X-ray luminosity estimate in
Cropper et al (1998): this due to the higher absorption originally
derived using the {\ros} PSPC spectrum. Figure \ref{pdot} implies that
it has reached $\sim4\times10^{35}$ \ergss in the past.

If the X-ray flux is driven by accretion and all of the accretion flow
was liberated as X-rays, then the required mass transfer rate would
exceed $7\times 10^{17}$ g s$^{-1}$ (or 1.2$\times10^{-8}$
\Msun/yr). This would place it at the upper end of the range of mass
transfer rates in known magnetic cataclysmic variables (Patterson
1994).  However, the UI model proposed by Wu et al (2002) is well
within the constraints of the observed X-ray luminosity, despite its
dependence on the orbital period of the system.  Although the maximum
power for a system with an orbital period of 569 sec, $M_{2}$=0.1\Msun
and a spin-orbit asynchronicity of 0.1 percent is less than
$\sim10^{34}$ \ergss (Wu et al 2002), an increase in the
asynchronicity to 1.0 percent would easily give rise to the observed
luminosity.

Unless the dipole axis was closely aligned with the spin axis of the
white dwarf, then this asynchronicity may be expected to have an
observational signature. For an asynchronicity of 1 percent, then that
would give rise to a `beat' period of $\sim$16 hrs. The {\xmm} X-ray
observations are separated by 6 days and have relatively short
durations so are not particularly suitable to search for such a
period. A dedicated `whole-Earth' type of observing campaign is the
best method to detect the signature of the asynchronous motion.
 
We note that an increase of the asynchronicity parameter would not
significantly reduce the total lifetime of the binary, as the primary
driver of the orbital evolution is gravitational radiation, in spite
of short term effects caused by spin-orbit coupling between the stars
and the orbital rotation. An increase or decrease in $\alpha$ would
alter the duration of the duty cycle of the unipolar-induction
process, and hence the brightness and discovery probability of the
system. 

\subsection{The X-ray spectrum of RX J1914+24}

The X-ray spectra of RX J1914+24 in \S \ref{spec} showed a soft
blackbody component plus a broad feature resembling an emission line
with a central energy close to $\sim$0.59keV.  We fitted the spectra
with a range of possible models, including a two-temperature blackbody
model and found that this did not give good fits. More detailed work
is needed to determine if irradiated white dwarf atmosphere models can
provide better fits (see Williams, King \& Brooker 1987). Observations
of the disc accreting double degenerate binaries, the AM CVn systems,
show that their optical spectra can show metal abundances very
different from Solar (eg Marsh, Horne \& Rosen 1991). The relative
abundances are affected as a result of the CNO process and mixing in
the common envelope phase of the binary. X-ray observations of AM CVn
stars also show this non-solar abundance, eg Strohmayer (2004c),
Ramsay et al, in prep. Fits made using an emission model of different
metal abundances for each element gave better fits than a
two-temperature blackbody model, but poorer than a blackbody plus
Gaussian line.

Using the blackbody plus Gaussian line, we find that the broad line is
brightest at X-ray maximum, implying that it arises from a region
close to the hot spot where most X-rays are emitted. The high
resolution RGS spectra, however, do not show any significant evidence
for distinct individual lines. This is in contrast to {\chan}
observations of the 10.3 min binary ES Cet (Strohmayer 2004b) which
shows tentative evidence for narrow emission lines at 0.47 and
0.89keV. It is also in contrast to {\xmm} observations of the double
degenerate AM CVn system GP Com (Strohmayer 2004c) which show narrow
emission lines of NVII, NVI, Ne X and Ne IX. The line centre of
0.59keV is close to the O VII photoionised line at 0.57keV which has
been detected in some intermediate polars (Mukai et al 2003). The
lines in these systems, however, have much narrower width.

\begin{figure}
\begin{center}
\setlength{\unitlength}{1cm}
\begin{picture}(8,5)
\put(0,0){\includegraphics{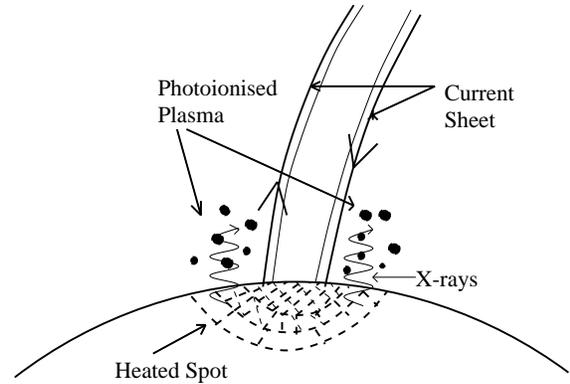}}
\end{picture}
\end{center}
\caption{A schematic diagram showing one possible mechanism for
producing the broad line seen near 0.6keV. The line may be broadened
due to the high velocities in the photo-ionized plasma.}
\label{cartoon}
\end{figure}

It is possible that the broad feature could be line emission from
material with large velocity dispersion. In the UI model the large
e.m.f. inside the magnetic flux tubes joining the two stars could
accelerate charged particles, thus causing difficulty in the
confinement of line emitting ionised atomic species right above the
X-ray emitting hot spot.  However, ionized particles could be present
in a region close to the hot spot, yet outside the magnetic flux tubes
that join the two stars (see the schematic illustration in Figure
\ref{cartoon}). Nevertheless, we argue that the feature is unlikely to
be due to a low-order harmonic hump of cyclotron emission, as it would
require a magnetic field strength $\sim 10^{11}$~G in the emission
region.  Such a field is much stronger than the strongest magnetic
fields measured in magnetic white dwarfs ($\sim10^{8-9}$ G, eg Barstow
et al 1995).

\subsection{The nature of RX J1914+24}

The true nature of RX J1914+24 is still unclear. Whilst the UI model
of Wu et al (2002) perhaps comes closest to predicting the observed
characteristics, it too has question marks. For instance, Barros et al
(2004) explored the set of binary parameters (masses, binary
inclination and the spin-magnetic axis offset) which could give rise
to the observed X-ray light curves of both RX J1914+24 and RX J0806+15
and concluded that only a very small set of parameter space could
work. This implies that either the UI model is unlikely to be
applicable to these systems or that the magnetic white dwarf has a
field structure more complex than a simple dipole.

The {\xmm} data presented here raise a number of issues, all of which
need further work to address them in detail:

\begin{enumerate}

\item RX J1914+24 has a highly unusual X-ray spectrum, with a peculiar
feature near 0.6keV. Grating spectra with higher signal to noise are
required and more theoretical work is needed to determine if the UI
model can produce such an X-ray spectrum.

\item We need to determine how system parameters, such as the
asynchronicity parameter $\alpha$, can produce the observed spin-up
rate and X-ray luminosuty.

\item Unless the spin and magnetic axes are very closely aligned, the
asynchronicity should give rise to a beat period.  More work is needed
observationally to search for such a period, and theoretically to
predict the amplitude of such a beat period.

\item Steeghs et al (2004) show an optical spectrum which has features
similar to that of a K star. To resolve the nature of these features
phase resolved spectroscopy is urgently required. If RX J1914+24 is a
triple system, then this will obviously have implications for the
systems evolutionary history.

\end{enumerate}

\section{acknowledgments}

We thank Andrew Willes for useful discussions and the anonymous
referee for useful suggestions which clarified aspects of the
text. This is work based on observations obtained with {\xmm}, an ESA
science mission with instruments and contributions directly funded by
ESA Member States and the USA (NASA). These observations were part of
the {\xmm} OM guaranteed time programme.

\end{document}